\documentclass[12pt]{article}
\usepackage[utf8]{inputenc}
\usepackage{graphicx}
\usepackage{authblk}
\usepackage{amsmath,amsfonts,amssymb}
\usepackage{booktabs} 
\usepackage[group-separator={,}]{siunitx}
\renewcommand{\Affilfont}{\footnotesize}

\usepackage{hyperref}
\usepackage{float}
\usepackage{wrapfig}
\usepackage{multicol}
\usepackage{subfig}

\usepackage{setspace}
\usepackage{colortbl}
\definecolor{light-yellow}{rgb}{1,1,0.8}
 
\usepackage[margin=0.75in]{geometry}
\usepackage[style=apa, url=false, backend=biber]{biblatex}
\DeclareSourcemap{
  \maps[datatype=bibtex, overwrite]{
    \map{
      \step[fieldset=editor, null]
      \step[fieldset=language, null]
      \step[fieldset=note, null]
    }
  }
}

\usepackage{hyperref}
\newbox{\myorcidaffilbox}
\sbox{\myorcidaffilbox}{\large\includegraphics[height=1.7ex]{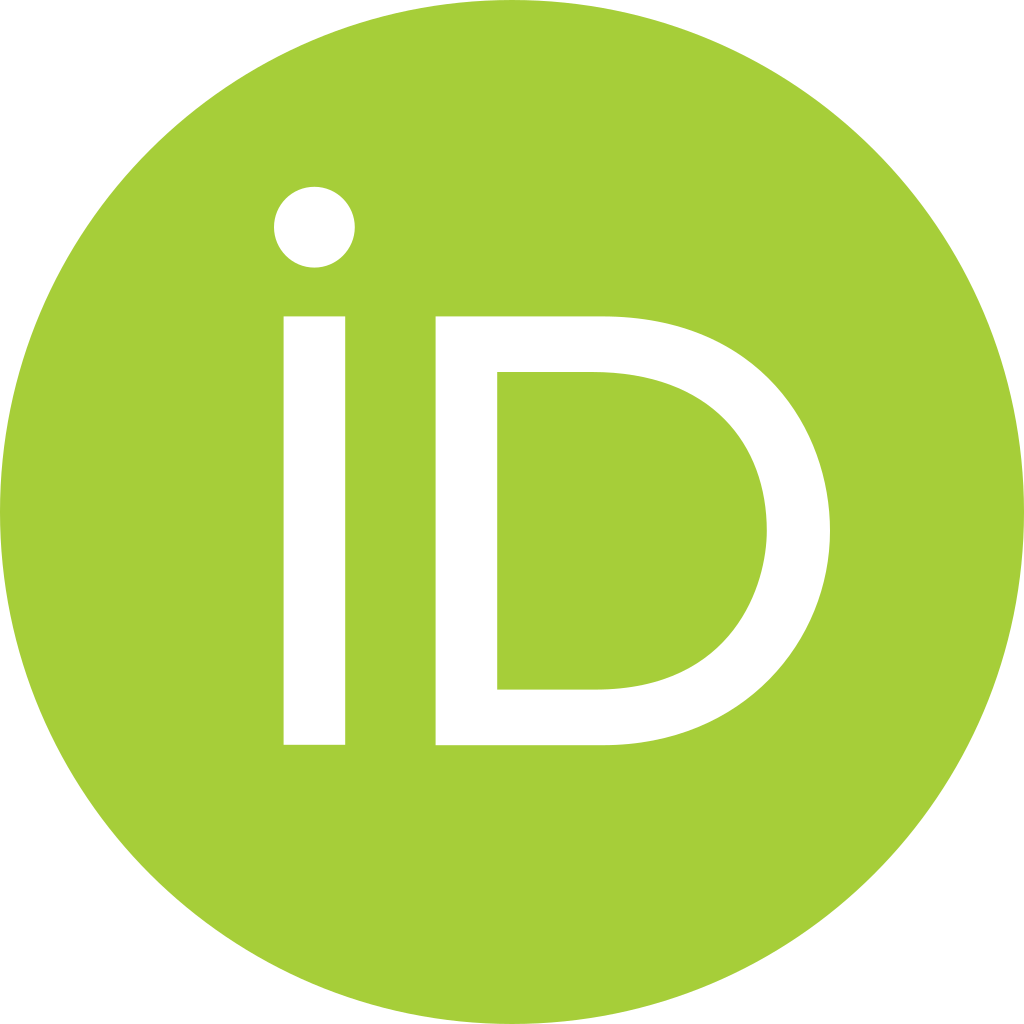}}
\newcommand{\orcidaffil}[1]{%
  \href{https://orcid.org/#1}{\usebox{\myorcidaffilbox}\,}}

\providecommand{\keywords}[1]
{
  \small	
  \textbf{\textit{Keywords:}} #1
}

\title{\vspace{-1.5cm}Representing the disciplinary structure of physics: a comparative evaluation of graph and text embedding methods}
\author[1]{Isabel Constantino \orcidaffil{0000-0002-4988-7146}}
\author[2]{Sadamori Kojaku \orcidaffil{0000-0002-9414-6814}}
\author[1]{Santo Fortunato \orcidaffil{0000-0002-9039-4730}}
\author[1,*]{Yong-Yeol Ahn \orcidaffil{0000-0002-4352-4301}}
\affil{Center for Complex Networks and Systems Research, Luddy School of Informatics, Computing, and Engineering, Indiana University, Bloomington}
\affil[2]{School of Systems Science and Industrial Engineering, Binghamton University}
\affil[*]{yyahn@iu.edu}
{
    \makeatletter
    \renewcommand\AB@affilsepx{: \protect\Affilfont}
    \makeatother

}
\date{January 7, 2025}

\begin{document}
\maketitle

\begin{abstract}
\noindent Recent advances in machine learning offer new ways to represent and study scholarly works and the space of knowledge.
Graph and text embeddings provide a convenient vector representation of scholarly works based on citations and text.
Yet, it is unclear whether their representations are consistent or provide different views of the structure of science.
Here, we compare graph and text embedding by testing their ability to capture the hierarchical structure of the Physics and Astronomy Classification Scheme (PACS) of papers published by the American Physical Society (APS). 
We also provide a qualitative comparison of the overall structure of the graph and text embeddings for reference.
We find that neural network-based methods outperform traditional methods, and graph embedding methods \textit{node2vec} and \textit{residual2vec} are better than other methods at capturing the PACS structure.
Our results call for further investigations into how different contexts of scientific papers are captured by different methods, and how we can combine and leverage such information in an interpretable manner.
\end{abstract}


\section{Introduction}
New discoveries and innovations build upon existing knowledge, often by combining previous knowledge together~\parencites{weitzmanRecombinantGrowth1998}{doi:10.1287/mnsc.47.1.117.10671}.
Subsequently, new knowledge can be used in many different domains in the future.
In this sense, each publication can largely be characterized by its \textit{references and citations} (i.e.,~the knowledge upon which the publication builds, and how future publications use its knowledge), and its \textit{content} (i.e., the new knowledge itself).
These two types of information have thus been used as the primary ways to study the development and impact of scientific knowledge.
For instance, the number of citations that a paper receives has been recognized as an important indicator of its importance since the onset of science studies~\parencites{garfieldCitationIndexingStudying1970}{garfield1978citation}.
Bibliometric coupling~\parencite{kesslerBibliographicCouplingScientific1963} and co-citation~\parencite{smallCocitationScientificLiterature1973} have also been used as crucial ways to glean into the contexts and clusters that papers form~\parencites{smallStructureScientificLiteratures1974}{griffithStructureScientificLiteratures1974}.
More recently, machine learning methods, particularly representation learning methods, have been increasingly used to capture the information stored in the texts as well as the citation network~\parencites{tshitoyanUnsupervisedWordEmbeddings2019}{pengNeuralEmbeddingsScholarly2021}.
Here, we use the corpus of physics papers from American Physical Society (APS) and its hierarchical classification system---Physics and Astronomy Classification Scheme (PACS) to evaluate how well existing text and graph embedding methods capture the hierarchical structure of physics. 

\subsection{What is embedding?}

Representation learning is a subfield of machine learning where feature representations are automatically learned from raw data.
These representations are then used for other tasks such as classification and prediction, or for further analysis of the data~\parencite{bengioRepresentationLearningReview2013}.
Embedding is a form of representation learning where dense, low-dimensional vectors (also known as \textit{embeddings}), are learned to represent discrete or high-dimensional data~\parencites{bengioNeuralProbabilisticLanguage2003}{morinHierarchicalProbabilisticNeural2005}.
Embedding aims to quantify the relationships between entries in the dataset, creating a meaningful vector space where similar data points are represented using similar vectors~\parencites{mikolovEfficientEstimationWord2013}{NIPS2013_9aa42b31}{mikolov-etal-2013-linguistic}.
While being similar to vector-space models of the past~\parencites{roweisNonlinearDimensionalityReduction2000}{belkinLaplacianEigenmapsDimensionality2003}, this new wave of embedding models distinguish themselves with the power of massive data and neural network's ability to learn meaningful representation from data~\parencite{mikolovEfficientEstimationWord2013}. 

Embedding has been widely used to represent words, documents, or networks.
In word (or document) embedding, the semantic relationships between words (or documents) are captured.
Most graph (or network) embedding maps the nodes of a network in a continuous space, providing a continuous representation of the relationships between nodes, as opposed to the discrete relationships represented by graph adjacency matrices~\parencites{10.1145/2623330.2623732}{groverNode2vecScalableFeature2016}.
The learned representations are then used for downstream tasks such as translation and link prediction.

\subsection{Quantifying and mapping relationships among scientific papers and fields}

There is a long history of mapping science, since the very beginning of science studies. 
Bibliometric coupling~\parencite{kesslerBibliographicCouplingScientific1963} and co-citation~\parencite{smallCocitationScientificLiterature1973} were identified as powerful ways to estimate the strength of relationship between individual publications, which is then used to map out scientific fields and key papers in each discipline~\parencites{smallStructureScientificLiteratures1974}{griffithStructureScientificLiteratures1974}{smallVisualizingScienceCitation1999}. 
Extending this line of research, Boyack, Klavans, and B\"orner created a map of scientific journals and disciplines, developed using clustering techniques that place similar (i.e.~in terms of cocitations and/or co-references) journals close to one another~\parencites{boyackMappingBackboneScience2005}{bornerDesignUpdateClassification2012}.
Further efforts on mapping science have been made using different scientific databases, and different measures of similarity~\parencites{boyackCreationHighlyDetailed2014}{vaneckVisualizingBibliometricNetworks2014}{guevaraResearchSpaceUsing2016}.
In addition, improvements in technology have made it possible to cluster individual documents. 
These are demonstrated in studies such as those by~\textcite{waltmanPrincipledMethodologyComparing2020} and~\textcite{boyackComparisonLargescaleScience2020a}, which compare various ways of clustering individual papers based on text and citation properties.
In particular, Waltman et al.~found that bibliographic coupling and BM25 (a text-based measure based on inverse document frequency) produced the most accurate clusters among the tested citation-based and text-based relatedness measures, respectively.
Meanwhile, Boyack and Klavans found that clusterings based on hybrid measures performed most accurately, and that contents of clusters varied more than accuracy, thereby suggesting that citation and text are complementary.

An important application of the similarity between pairs of papers is the measure of novelty in science or innovation.
In particular, the novelty of a paper or patent is measured in terms of the cocitations between each pair of references: a novel paper cites less frequently cocited reference pairs compared to a ``conventional'' or non-novel paper.
In both patents~\parencite{doi:10.1287/mnsc.47.1.117.10671} and scientific papers~\parencite{uzziAtypicalCombinationsScientific2013, wangBiasNoveltyScience2017}, works that reference novel or unusual combinations of ideas and other prior work are more likely to be highly cited, or highly ``disruptive''~\parencite{uzziAtypicalCombinationsScientific2013, wuLargeTeamsDevelop2019, linNewDirectionsScience2022}.
Furthermore, research citing novel combinations of ideas are more likely to experience delayed recognition, but are also more likely to be cited by other fields~\parencite{wangBiasNoveltyScience2017}.
Overall, mapping of scientific research, as well as the measurement of the distance between ideas, are crucial in understanding how scientists collectively create, develop, and combine ideas in the context of the ``space'' of knowledge.  

\subsection{Embedding's potential in science studies}

One caveat in using existing categorizations of scientific work, as many of the studies mentioned in the previous subsection have done, is that they tend to be rigid and change slowly over time, even when science rapidly develops.
In an effort to overcome this limitation, previous studies have applied methods in topic modeling to analyze trends and ``hot topics'' in science. 
In particular, Latent Dirichlet Allocation (LDA) is used to learn representations of publications as probability distributions over scientific topics, and representations of topics as probability distributions over words~\parencites{bleiLatentDirichletAllocation2003}{griffithsFindingScientificTopics2004}{bleiProbabilisticTopicModels2012}.
LDA has also been used to learn representations of authors' body of work as probability distributions over topics~\parencite{10.5555/1036843.1036902}.
However, one challenge with LDA and other topic modeling methods is the need to specify the number of topics in the corpus, which is not always known, or may change over time.
Moreover, topic modeling approaches documents as ``bags of words''; it ignores the order and context of words.
Although models such as BERTopic~\parencite{grootendorstBERTopicNeuralTopic2022} have addressed the issue of documents being represented through the bag-of-words approach, the topic representations are still generated in this manner.
In addition, BERTopic continues to assume that each document contains one topic, which is not true especially in classification systems where a paper may be part of more than one category.

By contrast, neural embedding methods such as \textit{word2vec}~\parencites{mikolovEfficientEstimationWord2013}{NIPS2013_9aa42b31} take into account the context within which words or entities appear, and it represents them into continuous space without the need to use categorization (whether provided or learned from the data).
These methods have been shown to be capable at meaningfully representing the space of scientific knowledge.
For instance, in a study by \textcite{chinazziMappingPhysicsResearch2019}, papers published by the American Physical Society are represented using their subject categories, creating a map of specialties of physics research.
This map is then used to analyze the production of physics research of various countries. 
In another study by \textcite{pengNeuralEmbeddingsScholarly2021}, journals in the Web of Science (WOS) are represented using their citation network, which are then used to learn analogy graphs of journals, and visualize the ``spectra'' of soft versus hard science.

Despite the increasing use of embedding methods in the study of scientific production and innovation, relatively little attention has been paid to the systematic comparison of two primary embedding approaches---content and graph embedding.
While previous studies~\parencites{zhangP2VLargescaleAcademic2019}{kozlowskiSemanticRelationalSpaces2021} have explored the use of various graph and text embedding methods to represent scientific publications, the methods were evaluated separately using different metrics.
In particular,~\textcite{zhangP2VLargescaleAcademic2019} propose a new method \textit{P2V} which utilizes the sampling methods used in \textit{word2vec} to learn vector representations of papers from various datasets.
They then test their method on tasks such as paper classification and similarity. 
Meanwhile,~\textcite{kozlowskiSemanticRelationalSpaces2021} explore existing methods in text embedding to capture the ``semantic space'' of science, and graph embedding to capture the ``relational space'', and combine these embeddings to study the development of science and technology studies.
In addition, neither of these studies compare their results to baseline non-embedding methods.

In this study, we compare popular graph and text embedding (specifically spectral and neural network based) methods in terms of how well the resulting embeddings capture the disciplinary structure of physics.
We focus on the field of physics, particularly papers published in the American Physical Society (APS), which, until recently, has used a hierarchical classification system called the Physics and Astronomy Classification Scheme (PACS).
We then conduct a series of experiments to evaluate the embeddings' ability to capture the hierarchical structure of PACS.

\section{Data and Methods}

\subsection{Data}
The APS provides a citation network of all articles published in its journals, as well as papers' metadata on authors, affiliations, titles, publication dates, etc.
We consider the citation network of all peer-reviewed scientific papers, reviews, and letters in all APS journals published up to the year 2010.
We remove articles such as announcements, comments and replies, errata, and retractions.
The resulting citation network consists of $452{,}096$ papers and $4{,}931{,}143$ citations.
We note that this includes papers published before 1979, which do not have PACS codes.
These are included in all embeddings; however, they are not considered in the evaluation. 

For the content embedding, although it would be ideal to use the full text of papers, this is usually not available to researchers.
Moreover, some more modern text embedding methods such as BERT~\parencite{devlinBERTPretrainingDeep2019} and its derivatives limit each input to the first 500--1000 ``tokens'' or word pieces.
Therefore, we consider the titles and abstracts of the papers in the citation network.
Because the data provided by the APS does not contain abstract information, we match entries in the APS to entries in the Web of Science using exact DOI matching.
Among the $358{,}478$ papers with exact DOI matches to the Web of Science up to 2010, $159{,}375$ have abstracts.
We limit the preprocessing of text data to removing HTML and MathML tags, as well as removing extraneous whitespace.

\subsection{Embedding} \label{sec:emb}
\subsubsection{Graph embedding}
We embed the citation network of the 452K papers by using graph embedding.
Graph embedding places one paper in a space in relation to other papers. 
There are two types of relations, i.e., references (out-going citations) and citations (in-coming citations), which provides semantically different views of a paper.
The references of a paper imply the knowledge relevant to the content of the paper, while the citations to a paper imply how the paper is utilized by the future papers.
We incorporate both aspects of relations into consideration by ignoring the edge directionality and producing the graph embeddings using the undirected citation network.

A traditional family of graph embedding methods is based on matrix factorization, with one of the most popular being Laplacian Eigenmap~\parencite{belkinLaplacianEigenmapsDimensionality2003}.
Laplacian Eigenmap is a spectral embedding method, which obtains an $m$-dimensional embedding by concatenating the second to the $(m+1)$-th largest eigenvectors of the symmetric normalized Laplacian $L = I - D^{-\frac{1}{2}}AD^{\frac{1}{2}}$, where I is the diagonal identity matrix, D is the diagonal degree matrix $D_{ii} = \sum_{j}A_{ij}$, and $A$ is the graph adjacency matrix.

The second graph embedding method is \textit{node2vec}~\parencite{groverNode2vecScalableFeature2016}. 
\textit{node2vec} is a direct application of the skip-gram negative sampling method (SGNS),  popularly known as \textit{word2vec}~\parencites{mikolovEfficientEstimationWord2013}{NIPS2013_9aa42b31}, a word embedding model that produces an embedding of words from given sentences.
\textit{node2vec} is adapted to network embedding by using a random walk, where each sentence is composed of nodes visited by a random walk. 
\textit{node2vec} allows the random walker to preferentially visit the previous nodes, or nodes that are not adjacent to previous nodes.
This bias is controlled by hyperparameters $p$ and $q$.
A value of $p=1$ and $q=1$ is equivalent to a uniform random walk.
For this study, we use $p=1$ and $q=1$, similar to previous studies evaluating graph embeddings~\parencites{guPrincipledApproachSelection2021}{dehghan-kooshkghaziEvaluatingNodeEmbeddings2022}{liuSTABLEIdentifyingMitigating2023}.

Finally, the third graph embedding method is \textit{residual2vec}~\parencite{kojaku2021residual2vec}.
\textit{residual2vec}, also a random walk-based embedding method, uses the bias removal mechanism of the negative sampling method used in \textit{node2vec} to remove any prescribed bias from the embedding.
Here, we consider several structural biases present in the citation network.
First, older papers are more likely to have a higher degree (i.e.~more citations) due to the time it has to accumulate citations. 
Second, papers tend to cite more recent references.
To address these biases, we use the configuration model as a null model for \textit{residual2vec}.
In other words, we remove the bias caused by each paper's degree from the embedding.
For all graph embeddings, we use dimension $d=128$, given that most real-world networks do not require a large number of dimensions, and that the embedding tends not to suffer much from overparametrization of the number of dimensions~\parencite{guPrincipledApproachSelection2021}.
Other parameters we use are: walk length $l = 80$, walks per node $r = 10$, context window size $k = 10$, and random walk restart probability $s = 0.01$.

\subsubsection{Text embedding}
We use three popular text embedding methods: \textit{doc2vec}~\parencite{leDistributedRepresentationsSentences2014}, SciBERT~\parencite{beltagySciBERTPretrainedLanguage2019}, and Sentence-BERT~\parencite{reimers-gurevych-2019-sentence}.
\textit{doc2vec} learns document representations by training them to predict words in the document, similar to how \textit{word2vec} learns word representations by predicting words within its context.
We specifically use the Paragraph Vector-Distributed Memory (PV-DM) model of \textit{doc2vec}, where the paragraph (or document) vector serves as a ``memory'' that provides a broader context.
For \textit{doc2vec}, we use dimension $d=128$, window size 5, and minimum word instance count 5.

The second text embedding method is SciBERT~\parencite{beltagySciBERTPretrainedLanguage2019}, an extension of BERT~\parencite{devlinBERTPretrainingDeep2019}.
BERT is a pretrained deep neural language model that is trained using a bidirectional transformer network, and can be fine-tuned for tasks such as classification and question answering.
SciBERT is a BERT model finetuned on a dataset of computer science and biomedical science papers.
We use the \texttt{allenai/scibert\_scivocab\_uncased} model hosted on the Huggingface Transformers library~\parencite{wolf-etal-2020-transformers}.
As BERT generates representations for word pieces or punctuation marks (also known as \textit{tokens}), some extra processing is required to obtain representations of ``sentences'' or documents.
The processing is as follows~\parencite{reimers-gurevych-2019-sentence}: (1) For each text input or document, obtain the special \texttt{CLS} token; (2) for each \texttt{CLS} token, obtain the vectors from the last 4 hidden layers of the neural network; and (3) sum up the 4 layers.

The last text embedding method is Sentence-BERT~\parencite{reimers-gurevych-2019-sentence}.
Sentence-BERT extends on BERT using a Siamese network architecture to obtain sentence embeddings that can be directly compared using measures such as cosine similarity.
The pretrained model used is \texttt{sentence-transformers/} \texttt{paraphrase-mpnet-base-v2}, also hosted by Huggingface.
As Sentence-BERT already results in an embedding for each piece of text, no further processing is required.
We note that both SciBERT and Sentence-BERT support a maximum of 512 word tokens; any input beyond 512 tokens is ignored.

\subsection{Evaluation}
\begin{figure}
    \centering
    \includegraphics[scale=0.25]{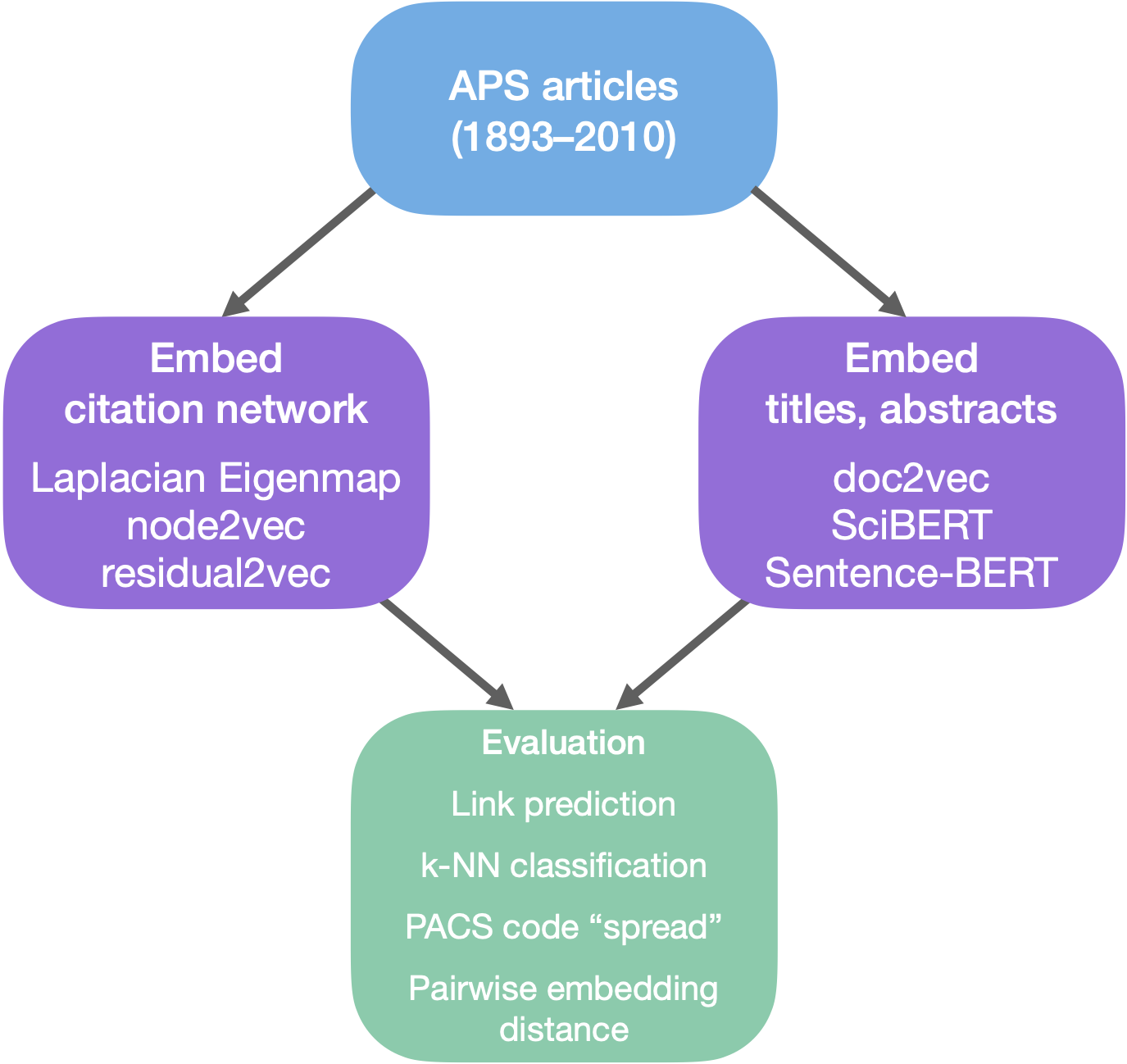}
    \caption{Embedding and evaluation framework for APS papers.}
    \label{fig:eval}
\end{figure}

What makes a good embedding? 
There are a multitude of ways in which the parameters of each embedding method can be tuned, and many other ways to judge ``similarity'' or distance'' between two embedded entities.
We summarize the embedding and evaluation framework in Figure~\ref{fig:eval}.

\subsubsection{About the PACS code}
Because we are interested in how each of the two main families of embedding methods can encode the hierarchical disciplinary structure of physics, we evaluate the embeddings based on how well they reflect an existing hierarchical categorization method, the Physics and Astronomy Classification Scheme (PACS)~\parencite{PhysicsAstronomyClassification2008}.
We consider each PACS code at each level.
At the highest level, there are 10 PACS codes (00--90) each corresponding to a general physics category such as particle physics, nuclear physics, etc.
Each succeeding level then corresponds to a more specific subtopic.
Below is an example of a six-digit PACS code: \textbf{61.30.Jf}.
The first level, \textbf{6}, or \textbf{60}, corresponds to ``the structural, mechanical, and thermal properties of condensed matter''.
The second level, \textbf{61}, then represents ``crystallography'' and ``solid or liquid structures''.
The third level, \textbf{61.30}, refers to ``liquid crystals''.
Finally, the fourth or lowest level, \textbf{61.30.Jf}, refers to ``defects in liquid crystals''.
These six-digit codes are assigned by authors at the paper level, and each paper may have more than one PACS code.

\subsubsection{Classification}
We first evaluate the embeddings' ability to capture the general clusters of physics research, by running a $k$-nearest neighbor (KNN) classification task to predict the level 1 (L1) PACS code, that is, its general physics classification.
KNN is a supervised machine learning algorithm where data is labeled according to the labels of its $k$ nearest neighbors, as calculated by a distance metric.
We first take the citation network consisting of papers published on or before the year 2010.
The true labels are determined as follows: for each paper, we take all full PACS codes, and take the first digit of each PACS code to get the L1 PACS code.
In case of duplicates, we keep the most frequently used L1 PACS codes (in case of ties, we select the last L1 PACS code, as sorted randomly).
Then, we split the papers into training and testing sets, with a ratio of 80\% to 20\%, respectively.
The train-test split step is performed for each embedding; hence, the training and testing sets are different for all embeddings.
The papers in the testing set are labeled with the L1 PACS code of the majority of its $k$ nearest neighbors (which are part of the training set), as measured by the cosine distance between the paper vectors in the embedding space.  
For this study, we implement KNN using the \textit{Faiss} library~\parencite{johnsonBillionscaleSimilaritySearch2019}, which provides a faster calculation of distance in large, high-dimensional datasets.
We repeat for the values $k \in \{2, 4, 8, 16, 32, 64, 128\}$.
The resulting classification is then evaluated using the micro-F1 score. 

\subsubsection{Evaluating hierarchical disciplinary structure}
We then evaluate whether the embeddings can capture the hierarchical structure of PACS.
We do this in two ways.
First, we check whether papers of more specific subcategories are closer to one another, compared to papers of general subcategories.
Intuitively, sets of papers of a same general topic ``cover'' more material than sets of papers of a more specific topic. 
We want to check whether each succeeding PACS level has smaller and smaller variability among its constituent papers.
We do this using a measure inspired by the radius of gyration (ROG), or the root mean square distance of an object's parts to its center of mass.
In our case, we measure the root mean square cosine distance of each vector, to the centroid or average of these vectors:
\begin{equation}
    \text{ROG}_C = \sqrt{\frac{1}{N} \sum_{\mathbf{x} \in C}{\Big( 1 - \cos(\mathbf{x}, \mathbf{\Bar{x}}) \Big)^2}}
      \label{eq:rog}
  \end{equation}
where $C$ is the PACS code of interest, $\mathbf{x}$ a vector representation of a paper in $C$, $\mathbf{\Bar{x}}$ the centroid of all vectors in $C$, and $\cos(\mathbf{a},\mathbf{b})$ the cosine similarity  between vectors $\mathbf{a}$ and $\mathbf{b}$.
We calculate the ROG of each PACS code at each of the 4 levels.
We then get the distribution of ROG values at each level.
To confirm whether each finer level has smaller or larger ROG values, we perform a one-sample Wilcoxon signed-rank text~\parencites{wilcoxonIndividualComparisonsRanking1945}{conover1999practical} to test the null hypothesis that the median ROG distribution of a PACS level is greater than or equal to the median ROG distribution of the next lower PACS level.

Second, we evaluate whether pairs of papers of the same PACS code are more likely to be closer to one another in the embedding space, compared to papers of different PACS codes.
We sample $15{,}000$ pairs of papers in each of the following categories:  (1) random, (2) different discipline or PACS code, (3) same discipline (L1 PACS code), and (4) same subdiscipline (L2 PACS code).
We then get the distribution of cosine distance between the pairs of papers in each subcategory.
We compare these distributions in two ways.
First, we compute the Jensen-Shannon divergence between each of the cosine distance distributions.
The Jensen-Shannon (JS) divergence is a similarity measure for two distributions~\parencite{endresNewMetricProbability2003}.
Unlike the popular KL divergence~\parencite{kullbackInformationSufficiency1951}, it is symmetric and has a finite range of $0$ to $1$.
However, The Jensen-Shannon divergence does not indicate direction; that is, whether the cosine distance of same-discipline pairs is the same of that of random pairs.
Hence, we also conduct one-sample t-tests to compare the distributions.
In particular, we test the null hypothesis that the mean distance between random pairs is less than or equal to than the mean distance between same-discipline pairs.

\subsubsection{Link prediction}
Finally, we evaluate whether the embeddings are useful for predicting links between papers and references, specifically, whether a paper cites or references a previous paper~\parencites{adamicFriendsNeighborsWeb2003}{liben-nowellLinkpredictionProblemSocial2007}{shibataLinkPredictionCitation2012}.
We first take the citation network of APS papers published up to the year 2010.
We then take the largest connected component of this citation network, then sample 50\% or approximately $149{,}000$ edges from papers published in 2010 (i.e.~edges corresponding to citations coming from papers in 2010).
This is the ``positive'' example set.
We then sampled the same number of paper pairs (such that one of the papers in each pair is from 2010) that have \textit{no} edge between them, to serve as the ``negative'' examples.
The ``positive'' examples or edges are removed from the citation network, and the resulting network is embedded using each of the embedding methods described in section~\ref{sec:emb}.
Then, for each embedding, we calculate the cosine similarity of the positive and negative examples.
In this case, positive examples, or pairs of papers linked by citation, should have a high cosine similarity, while negative examples should have low cosine similarity.
The results are evaluated using the ROC curve, used in binary classification: the ROC curve plots the false positive rate against the true positive rate, denoting how the true positive rate changes as the false positive rate threshold is increased.
The area under the ROC curve (AUC) is then calculated to summarize the results.
The AUC ranges from 0 to 1, where 1 represents a perfect classification, and an AUC of 0.5 represents a random classification.

\section{Results}
\begin{figure}[!htb]
        \centering
        \includegraphics[width=1\linewidth]{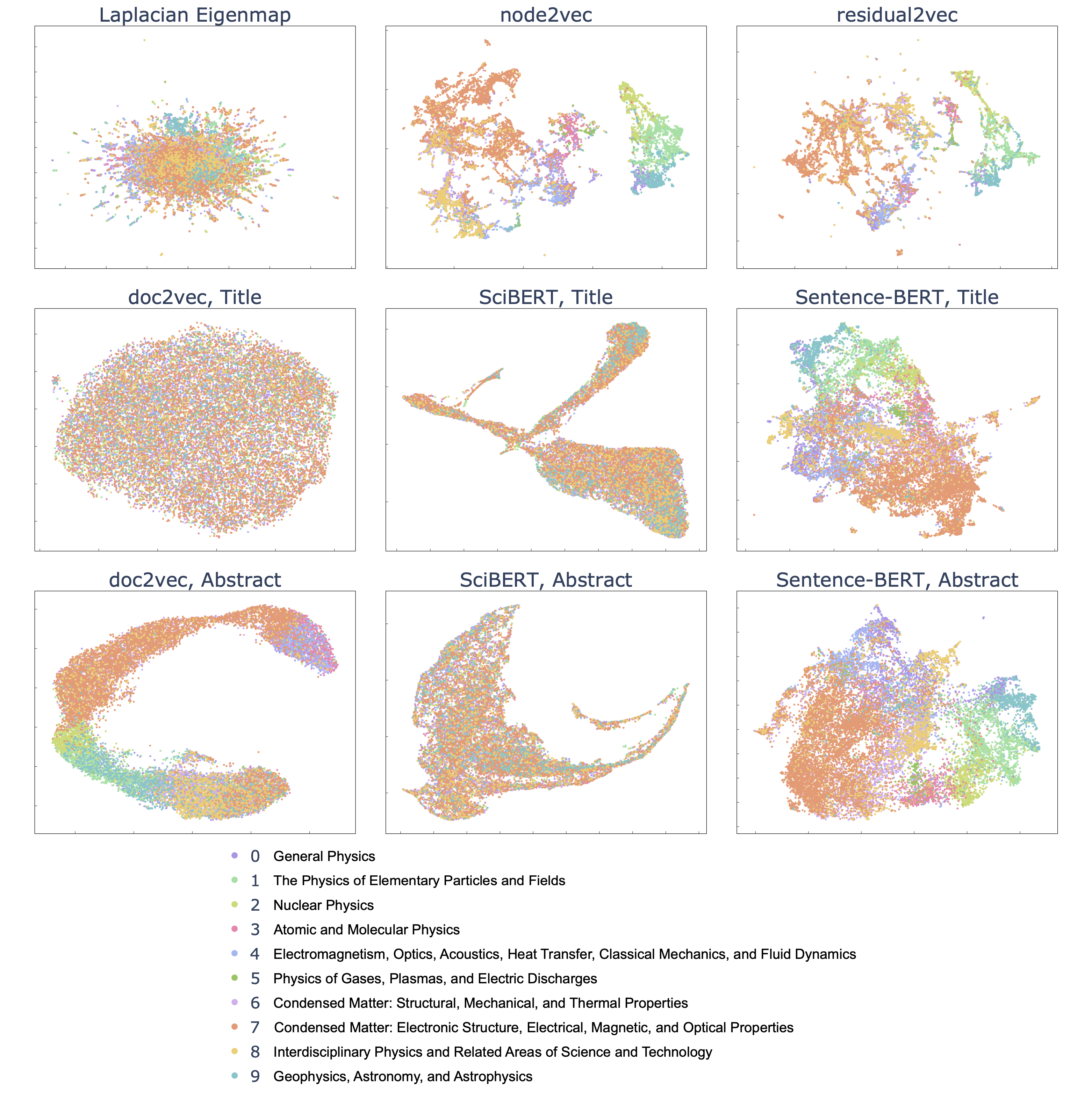}
        \caption{UMAP projections on a sample of the embeddings show that \textit{Sentence-BERT},\textit{node2vec}, and \textit{residual2vec} all follow the general clustering structure of physics research.}
        \label{fig:umap}
\end{figure}

Figure \ref{fig:umap} shows UMAP projections of each embedding, using a sample of 10,000 papers.
UMAP is a dimensionality reduction method known for preserving the local and global features of high-dimensional data~\parencite{mcinnes2020umapuniformmanifoldapproximation}. 
Each point is colored according to its level 1 PACS code.
Here, \textit{Sentence-BERT}—both title and abstract embeddings—\textit{node2vec}, and \textit{residual2vec} provide paper-level embeddings that follow the general clustering structure of research published in APS.
The Laplacian Eigenmap embedding also somewhat exhibits a general cluster structure, although it also presents many overlaps among the clusters. 

\begin{figure}[H]
    \centering
    \includegraphics[width=1\textwidth]{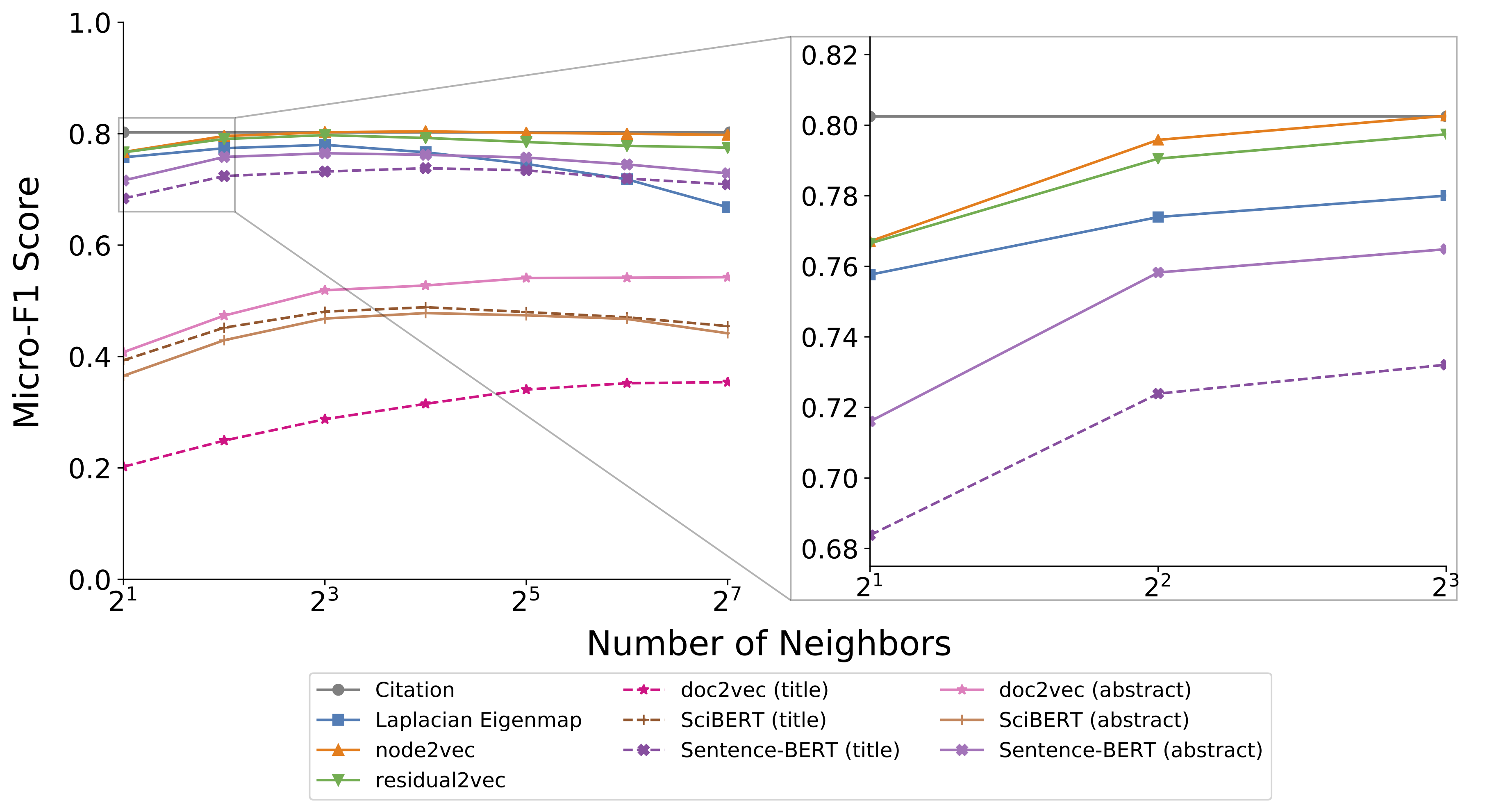}
    \caption{Classification of PACS by a k-nearest neighbor algorithm \textbf{(left)} with $k = 2$ to 128; and \textbf{(right; inset)} with micro-F1 score $> 0.68 $ and $k \in \{ 2,4,8 \}$. The graph embedding methods outperform all text  embedding methods.  Among the text embeddings, \textit{Sentence-BERT} performs best, though not as well as the graph embeddings. With \textit{doc2vec}, the abstract embedding results in improved classification performance compared to the title embedding.}
    \label{fig:knn}
\end{figure}

\clearpage
Figure \ref{fig:knn} shows the results for the $k$-nearest neighbor classification. 
Among the text embeddings, \textit{Sentence-BERT} has the highest micro-F1 score of approximately $0.75$, across multiple values of $k$. 
All text embeddings performed worse than the tested graph embeddings, which all resulted in micro-F1 scores of approximately $0.76$ to $0.81$.
In Laplacian Eigenmap, higher values of $k$ result in lower micro-F1 scores.
Finally, all embeddings except \textit{node2vec} performed worse than a baseline citation network-based classification which predicts a paper's level 1 PACS code using a majority vote among its references, although this may indicate a tendency for authors to self-assign PACS codes similar to those of their references.

\begin{figure}
        \centering
        \includegraphics[width=1\textwidth]{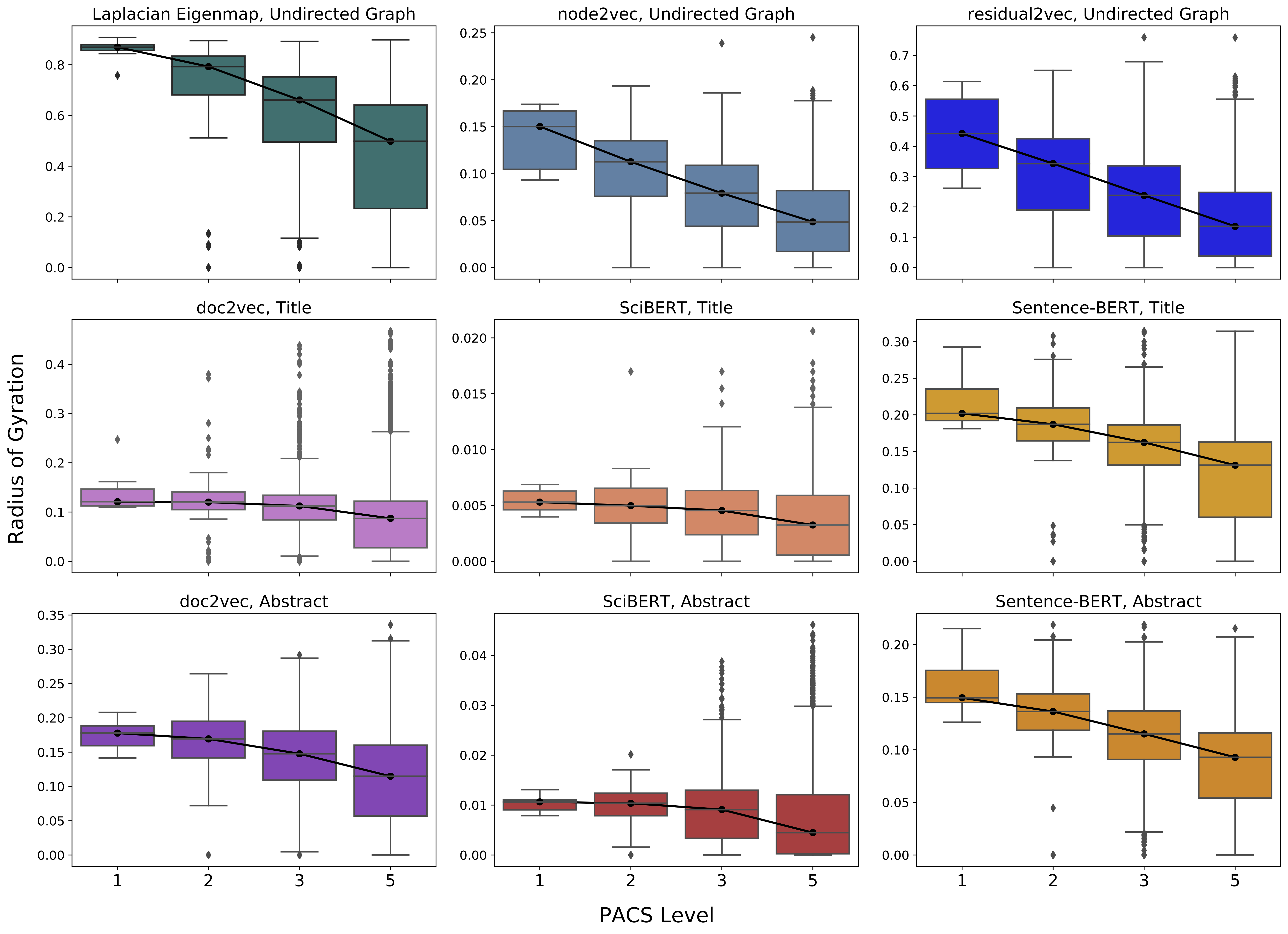}
        \caption{Box plots indicating the distributions of PACS code ROG show that a deeper PACS level (left to right) only results in a lower or more left-skewed ROG distribution in the \textit{Sentence-BERT}, Laplacian Eigenmap, \textit{node2vec}, and \textit{residual2vec} embeddings.}
        \label{fig:rog}
\end{figure}

\clearpage
Figure~\ref{fig:rog} shows the distributions of ROG at each PACS level, and Table~\ref{tab:rog_wtest} shows the results of the one-sample Wilcoxon test for the distributions.
For all tested graph embedding methods, as well as in \textit{Sentence-BERT}, the median ROG at each level is consistently lower than the median ROG of the preceding level.
Hence, for these embeddings, the more specific the subdiscipline of physics, the closer together its papers are in the embedding space.
Meanwhile, for \textit{doc2vec} and \textit{SciBERT}, the median ROG of the highest PACS level is lower than or equal to the median ROG of the second PACS level. 
This may be explained by the ``scale'' of these text embeddings, where \textit{all} points are relatively close to one another, thereby resulting in very small differences in ROG across the PACS levels.
By contrast, in the graph embeddings, the points are more spread out across the embedding space, which results in more dramatic decreases in ROG as the PACS level increases.
This is especially apparent in \textit{node2vec} and \textit{residual2vec}.
\begin{table}
    \centering
    \begin{tabular}{p{0.11\linewidth} | p{0.11\linewidth} | p{0.1\linewidth} | p{0.1\linewidth} |p{0.1\linewidth} | p{0.1\linewidth} |p{0.1\linewidth} | p{0.1\linewidth}}
    \toprule
         &               &   &  \# of samples & Median (current) & Median (previous) &  Wilcoxon statistic & p-value \\
Property & Embedding method & PACS level &               &                  &                   &                     &         \\
    \midrule
    Undirected Graph & Laplacian Eigenmap & 2 &            82 &           0.7928 &            0.8693 &                  43 &  0.00 \\
         &               & 3 &           941 &           0.6611 &            0.7928 &               17101 &  0.00 \\
         &               & 5 &          6434 &           0.4983 &            0.6611 &             2050166 &  0.00 \\
         & node2vec & 2 &            82 &           0.1127 &            0.1502 &                 156 &  0.00 \\
         &               & 3 &           941 &           0.0793 &            0.1127 &               53679 &  0.00 \\
         &               & 5 &          6434 &           0.0487 &            0.0793 &             3724942 &  0.00 \\
         & residual2vec & 2 &            82 &           0.3430 &            0.4419 &                 382 &  0.00 \\
         &               & 3 &           941 &           0.2381 &            0.3430 &               60733 &  0.00 \\
         &               & 5 &          6434 &           0.1360 &            0.2381 &             3723006 &  0.00 \\
    \midrule
    Title & \textbf{doc2vec} & \textbf{2} &            \textbf{82} &           \textbf{0.1202} &            \textbf{0.1210} &                \textbf{1668} &  \textbf{0.44} \\
         &               & 3 &           946 &           0.1123 &            0.1202 &              167476 &  0.00 \\
         &               & 5 &          6495 &           0.0871 &            0.1123 &             5446270 &  0.00 \\
         & SciBERT & 2 &            82 &           0.0050 &            0.0053 &                1339 &  0.0469 \\
         &               & 3 &           942 &           0.0045 &            0.0050 &              168165 &  0.00 \\
         &               & 5 &          6442 &           0.0033 &            0.0045 &             6409506 &  0.00 \\
         & Sentence-BERT & 2 &            82 &           0.1871 &            0.2020 &                 982 &  0.0004 \\
         &               & 3 &           942 &           0.1624 &            0.1871 &               76993 &  0.00 \\
         &               & 5 &          6442 &           0.1312 &            0.1624 &             3161325 &  0.00 \\
    \midrule
    Abstract & doc2vec & 2 &            76 &           0.1693 &            0.1778 &                1059 &  0.0182 \\
         &               & 3 &           844 &           0.1477 &            0.1693 &               91863 &  0.00 \\
         &               & 5 &          5418 &           0.1148 &            0.1477 &             3279790 &  0.00 \\
        & \textbf{SciBERT} & \textbf{2} &            \textbf{76} &           \textbf{0.0104} &            \textbf{0.0106} &                \textbf{1216} &  \textbf{0.10} \\
         &               & 3 &           844 &           0.0091 &            0.0104 &              136064 &  0.00 \\
         &               & 5 &          5418 &           0.0045 &            0.0091 &             4877620 &  0.00 \\
         & Sentence-BERT & 2 &            76 &           0.1364 &            0.1494 &                 883 &  0.0013 \\
         &               & 3 &           844 &           0.1151 &            0.1364 &               61858 &  0.00 \\
         &               & 5 &          5418 &           0.0929 &            0.1151 &             2434065 &  0.00 \\
    \bottomrule
    \end{tabular}
    \caption{One-sample Wilcoxon test results for comparing the median of each PACS level to the previous level (rows in bold text indicate null hypotheses that we failed to reject). This further shows that for the graph embeddings and \textit{Sentence-BERT}, the ROG decreases as the PACS level increases or becomes more specific. However, this more clearly shows that in \textit{doc2vec} and \textit{SciBERT}, the ROG is also decreasing in all but the most general PACS levels.}
    \label{tab:rog_wtest}
\end{table}

\begin{figure}
        \centering
        \includegraphics[width=1\textwidth]{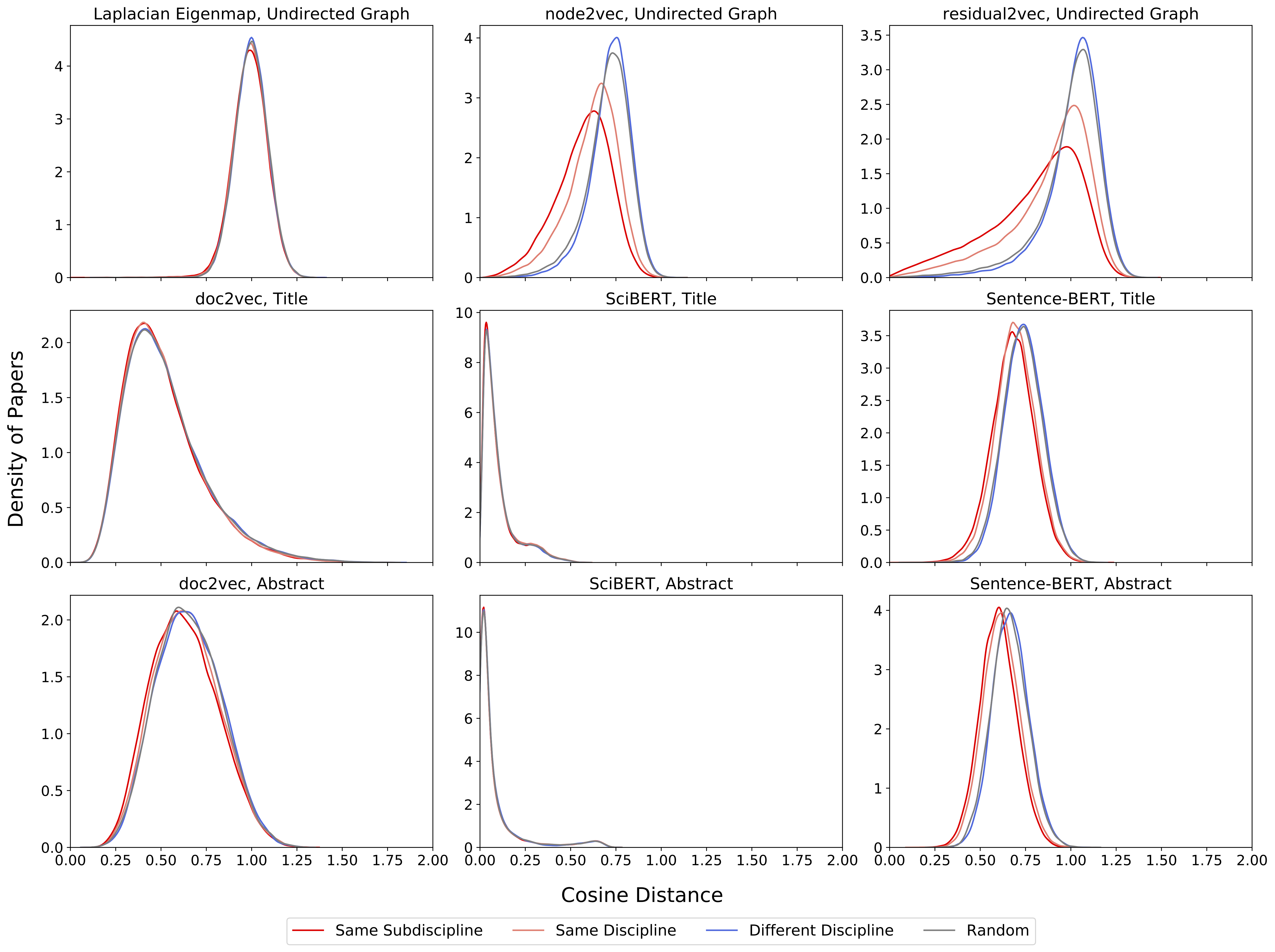}
        \caption{A comparison of embedding distance distribution between sampled paper pairs also show that \textit{Sentence-BERT} (title and abstract embeddings), \textit{node2vec}, and \textit{residual2vec} embeddings are more likely to embed papers of the same discipline closer to one another than random pairs of papers.}
        \label{fig:emb-dist}
\end{figure}

\begin{figure}
    \centering
    \includegraphics[width=1\textwidth]{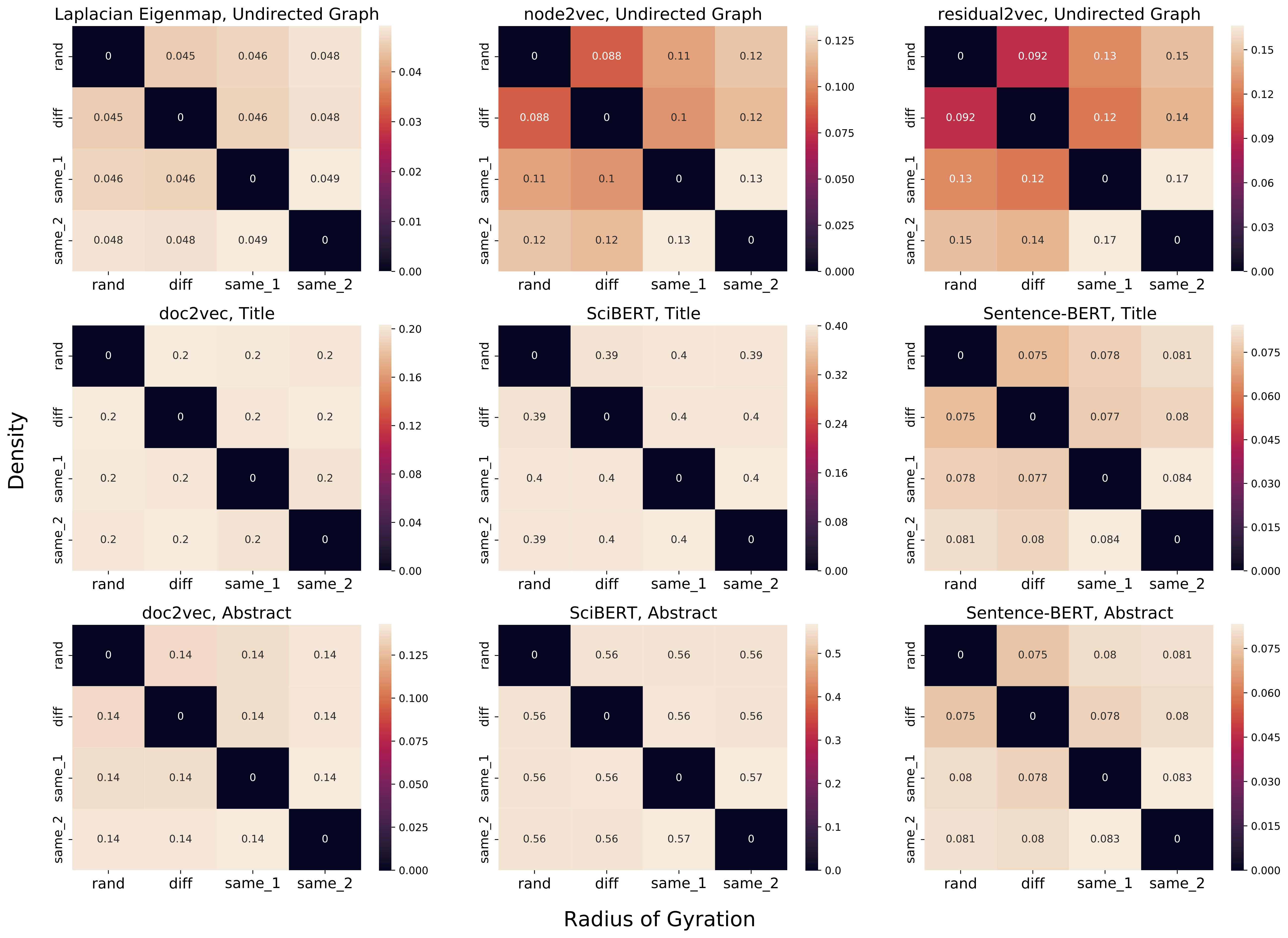}
    \caption{Jensen-Shannon (JS) distance matrices between cosine distance distributions of paper pairs. Here, \textit{node2vec} and \textit{residual2vec} have the most visible differences in JS distance across the multiple distributions. In particular, the JS distances are smallest along the diagonal, with the distances increasing for paper pairs with the same L1 vs. L2 PACS codes.}
    \label{fig:emb-dist-js}
\end{figure}

Figure~\ref{fig:emb-dist} shows the distributions of sampled paper pairs, Figure~\ref{fig:emb-dist-js} shows the Jensen-Shannon (JS) distance matrices for each embedding, and Table~\ref{tab:emb-dist-ttest} shows the results of the t-test to compare these embeddings' distance distributions.
In the representations generated using the graph embeddings and \textit{Sentence-BERT}, the mean cosine distance of random paper pairs is consistently greater than that of pairs with the same L1 or L2 PACS code. 
In addition, the JS distances between the distributions in the \textit{node2vec} and \textit{residual2vec} embeddings are increasing. 
That is, the JS distance between random pairs and pairs with different PACS codes is less than the JS distance between pairs with the same L1 and L2 PACS codes.
However, on \textit{doc2vec} embeddings on titles, we fail to reject the hypothesis that the mean cosine distance of pairs with different PACS codes is not equal to that of random pairs.
Moreover, there are no clear differences in Jensen-Shannon distance among the cosine distance distributions of random versus other paper pairs.
Meanwhile, on abstracts, although there are also no visible differences in Jensen-Shannon distance, we do reject the hypothesis that the mean cosine distance of random pairs is not equal to that of pairs with different PACS code.

\begin{table}
    \centering
    \begin{tabular}{p{0.11\linewidth} | p{0.12\linewidth} | p{0.5\linewidth} | p{0.09\linewidth} |p{0.08\linewidth} }
    \toprule
    Property & Embedding method & $H_a$ &  T          &  p-value       \\
    \midrule
    Undirected Graph & Laplacian Eigenmap & $\mu(\text{Different PACS}) \neq \mu(\text{Random})$ &     3.39 &  0.0007 \\
         &               & $\mu(\text{Same L1 PACS}) < \mu (\text{Different PACS})$ &   -15.31 &  0.00 \\
         &               & $\mu(\text{Same L2 PACS}) < \mu (\text{Same L1 PACS})$ &    -9.79 &  0.00 \\
         & node2vec & $\mu(\text{Different PACS}) \neq \mu(\text{Random})$ &    34.19 &  0.00 \\
         &               & $\mu(\text{Same L1 PACS}) < \mu (\text{Different PACS})$ &  -173.57 &  0.00 \\
         &               & $\mu(\text{Same L2 PACS}) < \mu (\text{Same L1 PACS})$ &   -88.51 &  0.00 \\
         & residual2vec & $\mu(\text{Different PACS}) \neq \mu(\text{Random})$ &    35.31 &  0.00 \\
         &               & $\mu(\text{Same L1 PACS}) < \mu (\text{Different PACS})$ &  -143.23 &  0.00 \\
         &               & $\mu(\text{Same L2 PACS}) < \mu (\text{Same L1 PACS})$ &   -82.22 &  0.00 \\
    \midrule
    Title & \textbf{doc2vec} & \textbf{$\mu(\text{Different PACS}) \neq \mu(\text{Random})$} &    \textbf{-1.13} &  \textbf{0.26} \\
         &               & $\mu(\text{Same L1 PACS}) < \mu (\text{Different PACS})$ &   -12.14 &  0.00 \\
         &               & $\mu(\text{Same L2 PACS}) < \mu (\text{Same L1 PACS})$ &    -3.28 &  0.0005 \\
        & \textbf{SciBERT} & \textbf{$\mu(\text{Different PACS}) \neq \mu(\text{Random})$} &     \textbf{0.02} &  \textbf{0.98} \\
        &               & \textbf{$\mu(\text{Same L1 PACS}) < \mu (\text{Different PACS})$} &     \textbf{4.74} &  \textbf{1.00} \\
         &               & $\mu(\text{Same L2 PACS}) < \mu (\text{Same L1 PACS})$ &   -10.22 &  0.00 \\
         & Sentence-BERT & $\mu(\text{Different PACS}) \neq \mu(\text{Random})$ &    18.09 &  0.00 \\
         &               & $\mu(\text{Same L1 PACS}) < \mu (\text{Different PACS})$ &  -104.80 &  0.00 \\
         &               & $\mu(\text{Same L2 PACS}) < \mu (\text{Same L1 PACS})$ &   -32.84 &  0.00 \\
    \midrule
    Abstract & doc2vec & $\mu(\text{Different PACS}) \neq \mu(\text{Random})$ &     2.11 &  0.0351 \\
         &               & $\mu(\text{Same L1 PACS}) < \mu (\text{Different PACS})$ &    -9.29 &  0.00 \\
         &               & $\mu(\text{Same L2 PACS}) < \mu (\text{Same L1 PACS})$ &    -9.99 &  0.00 \\
             & \textbf{SciBERT} & $\mu(\textbf{Different PACS}) \neq \mu(\textbf{Random})$ &    \textbf{-0.24} &  \textbf{0.81} \\
         &               & $\mu(\text{Same L1 PACS}) < \mu (\text{Different PACS})$ &    -1.98 &  0.02 \\
                &               & $\mu(\textbf{Same L2 PACS}) < \mu (\textbf{Same L1 PACS})$ &    \textbf{-0.27} &  \textbf{0.39} \\
         & Sentence-BERT & $\mu(\text{Different PACS}) \neq \mu(\text{Random})$ &     9.31 &  0.00 \\
         &               & $\mu(\text{Same L1 PACS}) < \mu (\text{Different PACS})$ &   -65.08 &  0.00 \\
         &               & $\mu(\text{Same L2 PACS}) < \mu (\text{Same L1 PACS})$ &   -17.34 &  0.00 \\
    \bottomrule
    \end{tabular}
    \caption{t-test results for comparing the mean cosine distance for each distribution. Rows in bold text indicate null hypotheses that we failed to reject.}
    \label{tab:emb-dist-ttest}
\end{table}

\begin{figure}
    \centering
    \includegraphics[width=0.8\textwidth]{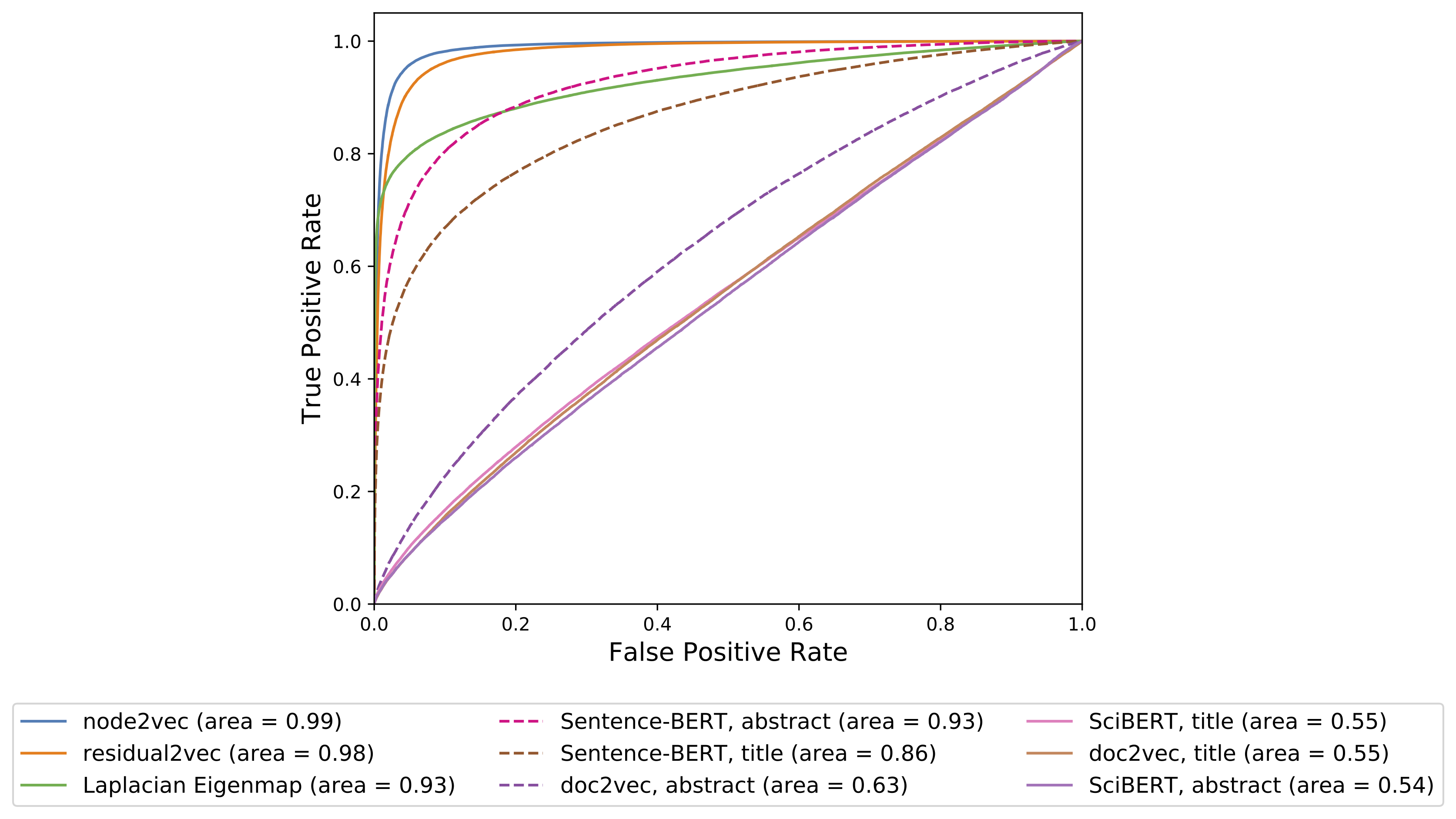}
    \caption{Link prediction results show that the graph embedding methods perform very well in link prediction, especially \textit{node2vec} and \textit{residual2vec}. Among the text embedding methods, \textit{Sentence-BERT} also performs relatively well.}
    \label{fig:linkpred}
\end{figure}

\clearpage
Figure~\ref{fig:linkpred} shows the ROC curves and AUC for the link prediction experiments.
The graph embedding methods have the highest performance for link prediction, as measured by the area under the ROC curve (AUC).
\textit{Sentence-BERT} performs best among the text embedding methods, but it does not perform as well as the random-walk based embedding methods \textit{node2vec} and \textit{residual2vec}.

\section{Discussion}
In this paper, we have compared text and graph embedding methods to assess their ability to encode the disciplinary structure of science. 
Our results suggest that the disciplinary structure of physics may be better encoded with graph embedding rather than content embedding, although text embedding methods perform remarkably well given how little information they use. 
We have found that graph embedding methods generally perform better, probably given their ability to utilize the rich information stored in the whole citation network structure.  
At the same time, the excellent performance of \textit{Sentence-BERT}, which is a readily-available off-the-shelf model, demonstrates that even just a few words in the paper title may provide a lot of information about what the paper is about. 
This may be a demonstration of the impressive power of large language models and the information that they implicitly encode in their weights.

Broadly, the fact that graph embedding performs slightly better may speak to the tradition of leveraging co-citation or bibliometric coupling to map science; the citation edges capture a lot of rich information about the context in which each publication exists, while full content data is scarce and difficult to consistently obtain for a large swath of scientific fields. 
This was also due to the fact that, before the advancement of deep learning and large language models, there was a scarcity of computational methods that can extract semantic information out of limited text (often just titles) we can access. 
We can safely assume that, especially given that text embedding with abstracts perform better than that with just the titles, the most text we can access, the more information we can, in principle, extract. 
However, there is a significant trade-off between the breadth of data coverage and the amount of textual information we can access; even abstracts are not readily available for a large portion of scientific publications, let alone full-text access.
Also note that, even if we have access to the full text of scholarly publications, there is still a challenge of distilling the \emph{right} information out of the long text. For instance, a publication can be characterized by many different aspects---the methods that it uses, the domain, research questions, implications, and so on---and this multi-faceted nature of scientific publication calls for nuanced ways to think about the `representation' of them. 
As science becomes more `open' and the access to full text becomes easier, this question would become more salient. 

There is also a bigger question about different types of information that text and citation network capture and how they can be leveraged differentially in science studies. 
The better understanding of embedding methods as well as the different types of information contained in different artifacts will let us be more mindful about the specific representation technique that we need to employ to address our research questions. 

Another point to consider is that the data and model details matter. 
For example, although \textit{SciBERT} is trained on scientific literature, it performs poorly. 
This may be due to the fact that it is trained with computer science and biomedical papers (field mismatch) or that the model training was not done well. 
Either way, it underlines the importance of model details and training data, as well as a potential lack of transferability in using large language models for science studies.

Our study has many limitations. 
Our study depends heavily on the Physics and Astronomy Classification Scheme (PACS), and as such, the evaluation methods we used are focused on evaluating the hierarchical structure of the relationships among the data and assumes the ``correctness'' of the code assignments. 
Given the fundamental challenge of classifying all of physics into a hierarchical structure, the hierarchical structure that PACS code captures is only an approximation of the true relational structure between the publications that we considered. 
Other potential tests could be the comparison of clusters among embeddings, such as the evaluation utilized by~\textcite{boyackComparisonLargescaleScience2020a}.
We have also focused more on comparing each embedding method individually, and have not considered whether differences in the availability of the text and network data may affect the results of the embedding comparison. 
For instance, the availability of abstracts may be skewed more towards recent data.
However, we would like to note that LLM-based text embedding methods are off-the-shelf and only operate on each individual paper's information without any fine-tuning on the dataset; in other words, the smaller sample size in our test does not affect their performance. 
In future embedding comparisons, more careful checking of data availability should be performed; alternatively, studies could be limited to datasets where both text and network data is available for each data point.
Although our evaluation may be a useful first step towards understanding the information captured in text and graph embedding, our study cannot say much about what kinds of different information are encoded differentially in each of the embedding and how they can be leveraged for science studies. 

Despite its limitations, our study demonstrates the crucial importance of thoroughly evaluating, comparing, and deeply understanding representation learning methods in scientific research. Careless model selection can lead to nonsensical or misleading results. For example, one cannot indiscriminately apply \textit{SciBERT} to a dataset of sociological texts merely because it is trained on ``scientific'' literature, without considering its training corpus and how well it captures meaning in the specific domain of interest. 
Our findings underscore that uncritical use of such models can produce highly misleading outcomes.

Through this work, we present a deeper dive on how embedding methods could be utilized in analyzing scientific research data, and provide more insight on the caveats and considerations when using these methods.
In the future, we may explore the variety of information captured by different embedding methods, as well as test embedding methods that use both text data and citation information. 
Embedding text and citation data provides many exciting opportunities to study how scientific knowledge is referenced, created, and shared.

\section*{Data and Code Availability}
The code used in the analysis is available at \url{https://github.com/sabsconstantino/emb-comp-aps}.
The American Physical Society dataset is available upon request at \url{https://journals.aps.org/datasets}. 
Abstract data from the Web of Science is proprietary and cannot be made available.

\section*{Author Contributions}
Isabel Constantino---Conceptualization, Data curation, Methodology, Software, Visualization, Writing---original draft, Writing---review and editing. Sadamori Kojaku---Methodology, Software, Writing---original draft, Writing---review and editing. Santo Fortunato---Conceptualization, Funding acquisition, Methodology, Writing---review and editing. Yong-Yeol Ahn---Conceptualization, Funding acquisition, Methodology, Supervision, Writing---original draft, Writing---review and editing.

\section*{Competing interests}
The authors declare that they have no competing interests.

\section*{Funding Information}
I.C., S.K., and Y.Y.A. are supported by the Air Force Office of Scientific Research under award number FA9550-19-1-0391. S.F. is supported by the National Science Foundation under award number 1927418, the Air Force Office of Scientific Research under award number FA9550-19-1-0354, and the National Institute of General Medical Sciences of the National Institutes of Health under award number R35GM146974.

\clearpage
\printbibliography

\end{document}